\documentclass{iau}
\usepackage{graphicx}

\title[eROSITA - Nearby Young Stars in X-rays] %% give here short title %%
{eROSITA - Nearby Young Stars in X-rays}

\author[J. Robrade]   %% give here short author list %%
{J. Robrade}

\affiliation{Hamburger Sternwarte, Germany\\email: {\tt jrobrade@hs.uni-hamburg.de}}

\pubyear{2015}
\volume{314}  %% insert here IAU Symposium No.
\pagerange{119--126}
\jname{Young Stars \& Planets Near the Sun}
\editors{J. H. Kastner, B. Stelzer, \& S. A. Metchev, eds.}
\begin{document}

\maketitle

\begin{abstract}
X-ray surveys are well suited to detect, identify and study young stars based on their high levels of magnetic activity and thus X-ray brightness.
The eROSITA instrument onboard the Spectrum-Roentgen-Gamma (SRG) satellite will perform an X-ray all-sky survey that surpasses existing data by a sensitivity increase of more than an order of magnitude.
The 4~yr survey is expected to detect more than half a million stars and stellar systems in X-rays.

%% add here a maximum of 10 keywords, to be taken form the file <Keywords.txt>
\end{abstract}

\firstsection % if your document starts with a section,
              % remove some space above using this command.
\section{Introduction}

The X-ray emission from cool stars originates from magnetic coronae that contain hot plasma at MK temperatures and
volume complete X-ray surveys have shown that coronae are ubiquitous around all types of cool stars \cite{schmitt97}.
Magnetic activity is driven by stellar dynamos with activity levels covering a range of $\log L_{\rm X}/L_{\rm bol} \approx -3 \dots -7$.
Activity is particularly strong in young stars due to their fast rotation and declines as stars
spin down during their lifetime through the loss of angular momentum via magnetized winds (see S. Matt, this Issue).
The efficiency of a stellar dynamo and thus the generation of magnetic activity depend not only 
on rotation, but also on the convective turnover time, thereby introducing a mass/spectral type dependent property with higher efficiencies at lower masses.
When combining both parameters in the Rossby number $Ro = P_{\rm rot}/\tau_{c}$, the dynamo efficiency can be uniformly described for solar and late-type stars.
The resulting stellar activity-rotation relation that has been studied in detail in X-ray astronomy, see e.g. \cite{piz03, wri11}.
Key features are a saturated regime with a constant activity level around $\log L_{\rm X}/L_{\rm bol} \approx -3$, an unsaturated regime with 
a power-law decline, $L_{\rm X}/L_{\rm bol} \propto Ro^{\beta}$ and a threshold at $Ro \approx 0.1-0.15$.
Summarized, at fast rotation X-ray brightness depends only on stellar luminosity and beyond a mass dependent rotation period it decreases with slower rotation.

The strong decline of coronal X-ray emission by several magnitudes over the stellar lifetime allows for an identification of stellar youth.
In addition, younger and more active stars are not only X-ray brighter, but also have hotter coronae and harder X-ray spectra.
Observations at multiple wavelengths have shown that in general emission from hotter plasma declines more steeply,
leading to a pronounced contrast in X-ray observations.
For example, the 'Sun in Time' project showed that early G stars are typically saturated for about 100\,Myr and faint by about two orders of magnitude at relevant X-ray energies
already during the first Gyr of their life, see e.g. \cite{gue97, rib05}.
Not only the average stellar activity level depends on age, but also
the average age at which a star of a given spectral type leaves the saturated X-ray regime.
Thereby a quite robust age-dating via X-ray observations is possible for ensembles of stars at ages above roughly 100~Myr, but
saturation effects, variability and intrinsic scatter, especially of rotation periods at stellar youth,
considerably restrict the applicability of this method for individual objects.
Nevertheless, a classification as 'young' often suffices, e.g. when a distinction between nearby young moving group members and older field stars matters.

\section{The eROSITA All-Sky Survey}

eROSITA (extended Roentgen Survey with an Imaging Telescope Array) is the primary instrument onboard the Spectrum-Roentgen-Gamma (SRG) satellite which will be launched around end 2016 and placed in an L2 orbit. 
It has been developed under the leadership of the Max Plank Institute for Extraterrestrial Physics (MPE, PI institute) and consists of seven co-aligned X-ray telescopes, each equipped with a CCD-type detector and a field of view (FOV) with a diameter of 1.03~deg. 
eROSITA will perform a 4~yr imaging all-sky survey at low to medium X-ray energies (0.3\,--\,10.0~keV).
The eRASS (eROSITA All-Sky Survey) is unprecedented, given its sensitivity, energy range, spectral and angular resolution as well as its eightfold sky coverage.

Important technical parameter of the eROSITA instrument are (on-axis/FOV average) a HEW of 15/28~arcsec at 1.5~keV, an eff. area of 2400/1400~cm$^{2}$ at 1.0~keV and an energy resolution (FWHM) of 60~eV at 0.3~keV and of 140~eV at 6~keV.
The all-sky survey scan paths follow mainly an ecliptical geometry and accumulate in total an average sky exposure depth of about 2.5~ks.
eROSITA/SRG is a joint German/Russian project and eROSITA data will be shared in equal parts between the partners; the respective priority areas are implemented as a sky-spit through the galactic center, the German side is the 'Galactic West' (gal. lon. 180\,--\,360~deg).
A description of the mission and its science objectives are given in the eROSITA Science Book, see \cite{mer12}.

\begin{figure}[t]
\begin{center}
\includegraphics[trim={0mm 10mm 0mm 4mm}, clip, width=125mm]{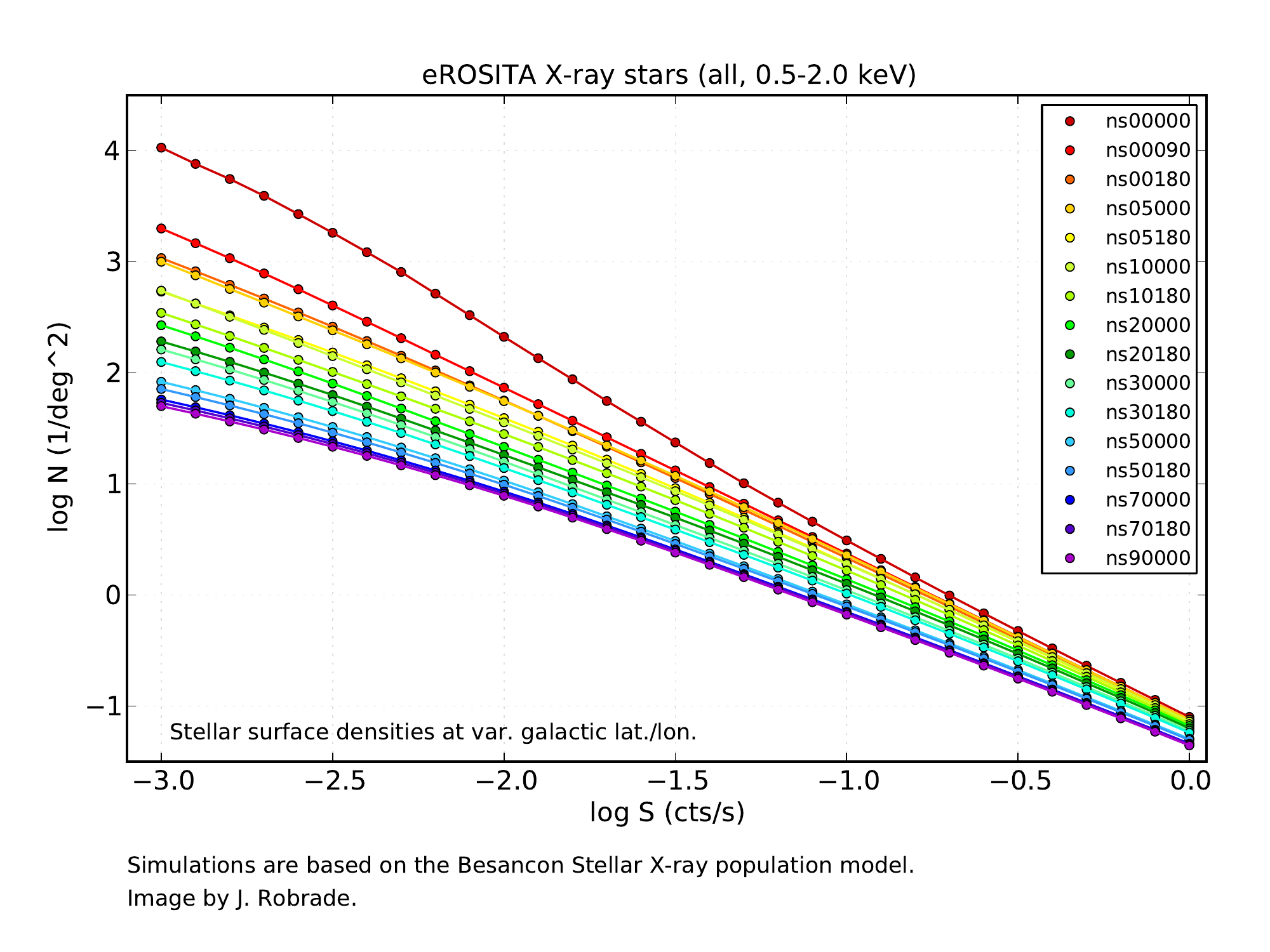} 
\caption{Simulated stellar densities in the eROSITA All-Sky Survey towards various galactic directions, the typical survey sensitivity is at $\log S \approx -1.9$\,cts\,s$^{-1}$.}
\label{xstars}
\end{center}
\end{figure}

With an average survey sensitivity of $F_{\rm X} \approx 1 \times 10^{-14}$~erg\,cm$^{-2}$\,s$^{-1}$
it is expected to detect about $0.7 \times 10^{6}$~stars, using simulations based on the Besancon Stellar X-ray population model \cite{gui96}.
As shown in Fig.\,\ref{xstars}, the typical stellar surface density is a strong function of galactic viewing direction; 
here the on average younger and X-ray brighter disk population further enhances the latitude contrast.
Separating the X-ray stars by mass/spectral type, late-type K and M~dwarfs dominate
given their X-ray brightness and large volume densities; separating by age, young stars (age $\lesssim 200$~Myr) contribute in total at a fraction of about 40\,\%
and start to dominate at low galactic latitudes. 
Naturally, they include the members of nearby young stellar populations, moving groups, clusters and associations.

The sensitivity increase of the eRASS by a factor of roughly 20 compared to the RASS (ROSAT All-Sky Survey) will help to overcome many issues related to
the brightness limits of the existing data and significantly enlarge the volume surveyed with sufficient depth at X-ray energies.
The deepened X-ray view of our galactic neighborhood and the multiple synergies have a high potential to set new levels in a variety of studies within the field of stellar astrophysics.

\begin{figure}[t]
\begin{center}
\includegraphics[width=100mm]{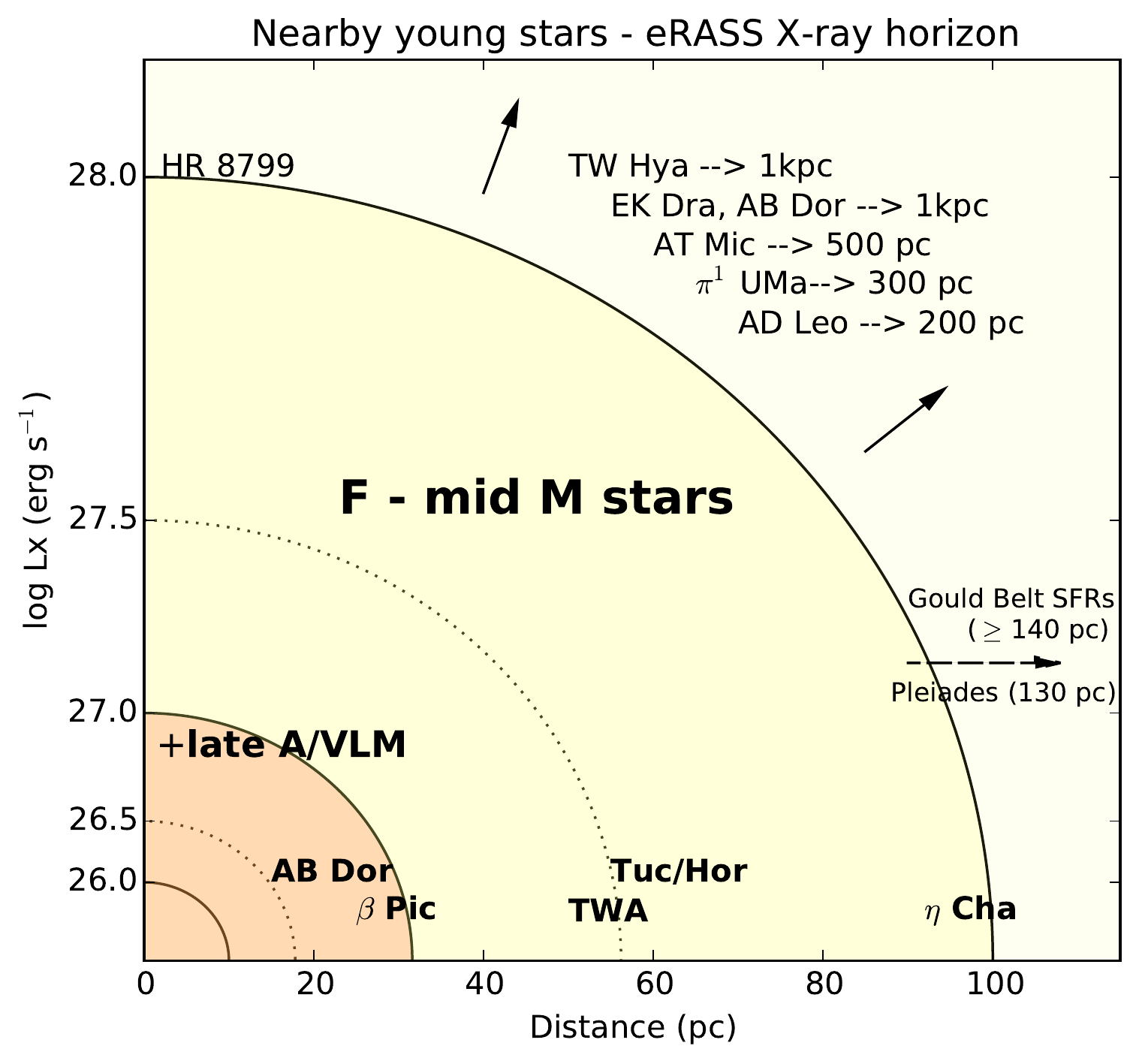} 
\caption{Sketch of the X-ray horizon of the eROSITA All-Sky Survey, average sky-sensitivity.}
\label{horizon}
\end{center}
\end{figure}

\section{Nearby young stars in the eRASS}

The 'local' X-ray sky is dominated by coronal emission from various types of cool stars and stellar systems, ranging from solar-type stars to very low mass stars and from pre-main sequence stars to older disk populations. Besides discovering new nearby young stars with the eRASS, their high X-ray fluxes also enable the investigation of coronal properties, variability and other characteristics. X-ray emission provides important diagnostics on magnetic activity phenomena and their evolution throughout the stellar lifetime, further it influences circumstellar disks and planetary systems. Overall, nearby young stellar populations are especially suited for X-ray studies.

The sketch shown in Fig.~\ref{horizon} depicts the typical eRASS sensitivity, when
following the rough scheme of this symposium by focussing on objects within 100~pc around the Sun that are younger than 200~Myr.
This region includes well studied moving groups and young associations like $\beta$\,Pic, AB~Dor, TWA, Tuc/Hor or $\eta$\,Cha.
For young stars within this volume the eRASS provides a virtually complete X-ray coverage of stellar targets with spectral types in the range of early~F to mid~M.
The eRASS sensitivity translates into a detection limit of $L_{\rm X} \approx 1 \times 10^{24} \times d^{2}$(pc)~erg\,s$^{-1}$ and
the X-ray horizons for exemplary young stars are indicated in the upper right corner. Taking again the 'young suns' example, stars like EK~Dra (G\,1.5, 100~Myr, $P_{\rm rot} = 2.7$~d, $L_{\rm X} \approx 10^{30}$~erg\,s$^{-1}$) can be detected up to a distance of about 1~kpc and stars like $\pi^{1}$\,UMa (G\,1.5, 300~Myr, $P_{\rm rot} = 4.7$~d, $L_{\rm X} \approx 10^{29}$~erg\,s$^{-1}$) up to roughly 300~pc.

In terms of activity levels, the flux limit transforms at a distance of 100~pc to sensitivities of $\log L_{\rm X}/L_{\rm bol} \gtrsim -6.3$ at spectral type early~F (5 L$_{\odot}$) and
$\log L_{\rm X}/L_{\rm bol} \gtrsim -3.3$ at mid~M (0.005 L$_{\odot}$).
This allows for the detection of virtually all so far unrecognized nearby young stars.
In addition, the sensitivity of the eRASS is sufficient to detect a large fraction of the members of moderately older (UMa cluster, Hyades) or more distant (Pleiades) groups,
opening the way to a more complete sampling of many well characterized, coeval stellar populations at ages of up to 1~Gyr.
By doing so, the eRASS sample broadens the stellar parameter space suitable for a systematic investigation of the magnetic activity evolution,
the activity-rotation-age relation and coronal properties, utilizing well defined stellar ensembles at different ages.
eROSITA will also detect X-ray emission from late A and late M stars, but towards higher masses the increasingly thinner convective zones
create weaker dynamos (low $L_{\rm X}/L_{\rm bol}$) and towards lower masses stars become fainter (low $L_{\rm bol}$).
Therefore stars at the hot and at the cool end of the magnetic activity range are intrinsically X-ray fainter and correspondingly their X-ray horizon will be closer.

In the regime of intermediate mass stars, the young HAeBe stars \cite{ste09} and the magnetic ApBp stars \cite{rob11} are prime targets for X-ray surveys. 
Both types of stars show diverse X-ray phenomena and a broad range of X-ray luminosities was observed in previously studied objects. The eRASS is promising to shed light on the general X-ray properties of the respective class. 

Going to sub-stellar objects, the most massive, youngest and hottest Brown Dwarfs \cite{wil14} are the most promising ones and X-ray detections will likely be limited to M-type Brown Dwarfs
that are either very nearby or in star forming regions.

\section{Star forming regions}

Active star formation is absent in the 100~pc solar neighborhood. However, several nearby groups like the TW~Hya Association ($d \approx 50$~pc, age $\sim$\,10~Myr) or the $\eta$~Cha cluster ($d\approx 97$~pc, age 6\,--\,8 Myr)
still contain a few classical T~Tauri stars (CTTS), i.e. young stars showing strong accretion signatures and a circumstellar disk.
Significant amounts of molecular gas or dust associated to these groups is virtually absent and
many of these older groups have already started to disperse, underlining the value of wide-field X-ray surveys for identifying and studying possible members.

The TWA and $\eta$~Cha cluster are thought to belong to the Sco-Cen OB association, which includes also active star forming regions like Chamaeleon, Lupus or Rho Oph. 
These are part of the Gould Belt, a nearby starburst region that contains further low-mass SFRs (star forming region) like Taurus-Aurigae and with Orion it harbors the closest active massive SFR. Massive stars are a different topic in X-ray studies, but of strong interest due to their importance for galactic evolution and chemical enrichment.

All classes of pre-main sequence stars are known to be X-ray bright and SFRs will be of prime interest to study the youngest stellar populations.
In contrast to the situation for young field stars, larger amounts of interstellar and circumstellar material are typically present and X-ray absorption becomes important.
The coverage of X-rays up to 10~keV, providing better sensitivity for embedded sources and the good positional accuracy are here further advantages of the eRASS data. 
Depending on distance and line-of-sight absorption, the evolutionary sequence from young CTTS over the virtually diskless WTTS (weak-line T~Tauri stars) 
to post~T~Tauri/ZAMS (zero-age main sequence) stars will be covered for a large range of stellar masses in many of these regions. In star formation studies, the eRASS
complements very well to deeper pointed observations with Chandra or XMM-Newton, that typically focus on the cores of SFRs.

\section{Beyond detections}

\begin{figure}[t]
\begin{center}
\includegraphics[width=70mm, angle=-90]{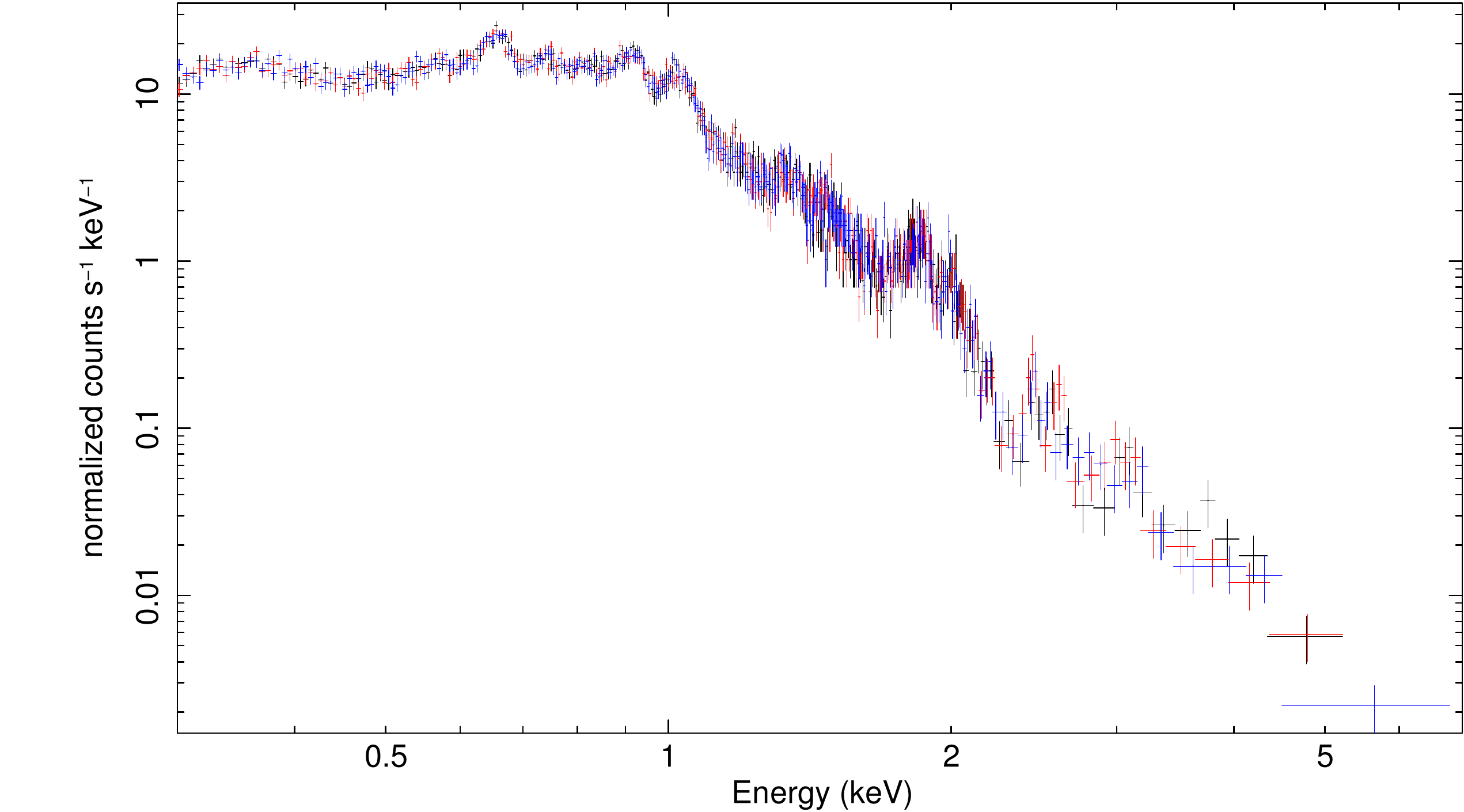} 
\caption{Simulated X-ray spectrum of the young M dwarf binary AT Mic, 2 ks survey exposure (eRASS/3 simulation runs).}
\label{specfig}
\end{center}
\end{figure}

Beside X-ray detections and fluxes, the eRASS will obtain spectra and light curves for all X-ray brighter sources and
many of these will be nearby young stars. Several thousands of stars have more than 1000 detected X-ray photons and several ten thousand more than 200 photons.
Fig.\,\ref{specfig} shows a simulated eRASS spectrum of AT~Mic, a mid M-dwarf binary that belongs to the $\beta$~Pic moving group.
The spectral quality is with about 25000 counts high enough to study the coronal plasma temperature distribution, elemental abundances or line-of-sight absorption;
even time-resolved spectroscopy becomes possible. 
If sample size is more important, lower quality data can be used to perform simple spectral modelling or derive the spectral hardness of the detected photons,
thereby putting constraints on coronal parameters or the classification of the respective object.

Variability is another important issue in stellar X-ray astronomy and many stars show distinct brightness variations.
The eRASS allows to investigate the time domain for many targets in greater detail and naturally provides X-ray measurements on several timescales. 
During the build up of the 4~yr survey, each source is visited at least eight times with a 0.5~yr cadence;
in addition the scanning law of the satellite provides several FOV passages during each visit with a cadence of 4~h.
Thereby the eRASS provides for the first time large sets of short- and long-term X-ray time series with a single instrument. 

Among the science themes to be addressed with time resolved data are the characterization of
typical X-ray activity levels and their long-term stability, coronal variability and its amplitudes, flare statistics and X-ray activity cycles.
Further, strong X-ray flares are quite common in a variety of stars, enabling the detection of additional sources with too low quiescent flux.
The X-ray spectra allow to address, among others, topics related to coronal physics, magnetic activity evolution
or the high-energy irradiance of circumstellar disks, exoplanets and their atmospheres.

\section{Synergies}

While the eROSITA data alone is already impressive, its scientific value is greatly enhanced when combining the X-ray sources and their properties with other suitable datasets,
most importantly to obtain identifications and distances as well as classifications and characterizations of their counterparts.
In this respect the optical brightness of the stellar content within the eROSITA survey is a great advantage. 
Most eRASS stars will have magnitudes of $V \lesssim 15$~mag and Gaia (see A. Sozetti, this Issue) will provide accurate positions, distances, 3-D space motions and basic stellar parameters;
the remaining fraction typically has at least partial or less accurate Gaia parameters.

In addition, a wealth of very useful multi-wavelength data from IR to UV wavelength exists, often with large area or all-sky coverage and several future wide-area surveys are planned.
Besides the numerous surveys in the optical regime, both photometric and spectroscopic (see K. Covey; S. Martell, this Issue), a 
virtually complete all-sky coverage is e.g provided by the NIR/IR surveys 2MASS and WISE.
Infrared observations are also an important supplement to optical data, when identifying very red or high-extinction sources.
In X-rays itself, the RASS and several mission archives enable the study of X-ray variability on timescales of many years up to decades.

Finally, large catalogs and data bases with rotation measurements or related activity indicators can also be combined with the X-ray data and dedicated follow-up observations for outstanding sources are envisaged.
Likewise nearby star searches like RECONS or SUPERBLINK and object class, planet host or young nearby moving group membership catalogs are just a few
examples of valuable complementary data to put the X-ray stars in a broader astrophysical context.

% \section{Conclusions}
% 
% The future is bright.

\bigskip\noindent
{\bf Acknowledgements}

I acknowledge support from DLR under the grant 50QR0803.

\begin{discussion}

\discuss{G. Hussain}{Will there be public data releases?}

\discuss{J. Robrade}{There will be publicly available eRASS data products/catalogs. Speaking for the German side, it is planned to have a final as well as intermediate
data releases after respective proprietary periods of about two years. If you are interested in collaborating, give us a note; the Hamburg Observatory is leading the eROSITA Stars WG.}

\end{discussion}


\begin{thebibliography}{}

\bibitem[ G{\"u}del et al. (1997)]{gue97}
{G{\"u}del, M. Guinan, E. \& Skinner, S.} 1997, 
\textit{ApJ}, 483, 947

\bibitem[(Guillout et al., 1996)]{gui96}
{Guillout, P., Haywood, M., Motch, C. \& Robin, A.C.} 1996, 
\textit{A\&A}, 316, 89

\bibitem[Merloni \etal\ (2012)]{mer12}
{Merloni, A., Predehl, P., Becker, W. et al.} 2012,
\textit{arXiv1209.3114M} 


\bibitem[Pizzolato et al. (2003)]{piz03}
{Pizzolato, N., Maggio, A., Micela, G., Sciortino, S., Ventura, P.} 2003, 
\textit{A\&A}, 397, 147

\bibitem[Ribas et al. (2005)]{rib05}
{Ribas, I., Guinan, E., G{\"u}del, M. \& Audard, M.} 2005, 
\textit{ApJ}, 622, 680

\bibitem[(Robrade \& Schmitt, 2011)]{rob11}
{Robrade, J. \& Schmitt, J.H.M.M.} 2011, 
\textit{A\&A}, 531A, 58

\bibitem[(Schmitt, 1997)]{schmitt97}
{Schmitt, J.H.M.M.} 1997, 
\textit{A\&A}, 318, 215

\bibitem[(Stelzer et al., 2009)]{ste09}
{Stelzer, B. Robrade, J., Schmitt, J.H.M.M. \& Bouvier, J.} 2009, 
\textit{A\&A}, 493, 1109

\bibitem[(Williams et al., 2014)]{wil14}
{Williams, P.K.G, Cook, B.A., Berger, E.} 2014, 
\textit{ApJ}, 785, 9

\bibitem[Wright et al. (2011)]{wri11}
{Wright, N.J., Drake, J.J., Mamajek, E.E. \& Henry, G.W.} 2011, 
\textit{ApJ}, 743, 48


\end{thebibliography}
\end{document}